\documentclass[twoside]{article}
\usepackage{Proc_NTSE,amsmath,amssymb,bbm,bm}

\begin{document}
\thispagestyle{plain}

\newcommand{\kev}{\,\mathrm{keV}}
\newcommand{\mev}{\,\mathrm{MeV}}
\newcommand{\gev}{\,\mathrm{GeV}}
\newcommand{\gevi}{\,\mathrm{GeV}^{-1}}
\newcommand{\fm}{\,\mathrm{fm}}
\newcommand{\fmi}{\,\mathrm{fm}^{-1}}
\newcommand{\fmiq}{\,\mathrm{fm}^{-3}}
\newcommand{\kf}{k_\text{F}}

\begin{center}
{\Large \bf \strut
Neutron matter with chiral EFT interactions: \\
Perturbative and first QMC calculations
\strut}\\
\vspace{10mm}
{\large \bf 
I.~Tews$^{a,b}$, T.~Kr\"uger$^{a,b}$, A.~Gezerlis$^{c,a,b}$, 
K.~Hebeler$^{a,b}$ and  A.~Schwenk$^{b,a}$}
\end{center}

\noindent{
\small $^a$\it Institut f\"ur Kernphysik, 
Technische Universit\"at Darmstadt, 
64289 Darmstadt, Germany} \\
{\small $^b$\it ExtreMe Matter Institute EMMI, 
GSI Helmholtzzentrum f\"ur Schwerionenforschung GmbH, 
64291 Darmstadt, Germany} \\
{\small $^c$\it Department of Physics, University of Guelph, 
Guelph, Ontario N1G 2W1, Canada}

\markboth{I. Tews, T. Kr\"uger, A. Gezerlis, K. Hebeler and A. Schwenk}
{Neutron matter with chiral EFT interactions} 

\begin{abstract}
Neutron matter presents a unique system in chiral effective field
theory~(EFT), because all many-body forces among neutrons are
predicted to next-to-next-to-next-to-leading order (N$^3$LO). We
discuss perturbative and first Quantum Monte Carlo (QMC) calculations
of neutron matter with chiral EFT interactions and their astrophysical
impact for the equation of state and neutron stars.
\\[\baselineskip] {\bf Keywords:} {\it Chiral EFT; three-body forces;
QMC; neutron matter; neutron stars}
\end{abstract}

\section{Chiral EFT and many-body forces}

Chiral EFT describes the interactions between nucleons at momentum
scales of the order of the pion mass $Q\sim m_{\pi}$ based on the
symmetries of QCD~\cite{RMP,EMRept}. The resulting nuclear forces are
organized in a systematic expansion in powers of $Q/\Lambda_\text{b}$,
where $\Lambda_\text{b}\sim 500 \mev$ denotes the breakdown scale,
leading to a typical expansion parameter $Q/\Lambda_{\text{b}} \sim
1/3$ for nuclei. At a given order this includes contributions from
one- or multi-pion exchanges that govern the long- and
intermediate-range parts and from short-range contact
interactions. The short-range couplings are fit to few-body data and
thus capture all short-range effects relevant at low energies.

In particular, chiral EFT provides a systematic basis to investigate
many-body forces and their impact on few- and many-body
systems~\cite{RMP3N}. In addition, it is possible to estimate
theoretical uncertainties in the EFT. An important feature of chiral
EFT is the consistency of two-nucleon (NN) and three-nucleon (3N)
interactions. This predicts the two-pion-exchange parts of the leading
(N$^2$LO) 3N forces, leaving only two low-energy couplings $c_D$,
$c_E$ that encode pion interactions with short-range NN pairs and
short-range three-body physics. At the next-order, all many-body
interactions are predicted parameter-free with many new
structures~\cite{RMP}. This makes the application of N$^3$LO 3N and 4N
forces very exciting. This is especially the case, because 3N forces
have been found to be key for neutron matter~\cite{nm} and for
neutron-rich nuclei~\cite{RMP3N,Oxygen}, see, e.g., the recent work on
the calcium isotopes~\cite{Calcium,CCCa,TITAN,pairing,ISOLDE}.

\section{Neutron matter from chiral EFT interactions}

The physics of neutron matter ranges from universal properties at low
densities~\cite{dEFT,GC} to the densest matter in neutron stars. For
neutrons, the $c_D, c_E$ parts of N$^2$LO 3N forces do not contribute
due to the Pauli principle and the pion coupling to the nucleon spin
(also the $c_4$ two-pion-exchange part does not contribute due to the
isospin structure)~\cite{nm}. Therefore, all three- and four-neutron
forces are predicted to N$^3$LO. To study these, we recently presented
the first calculation of the neutron-matter energy that includes all
NN, 3N, and 4N interactions consistently to
N$^3$LO~\cite{N3LO,longN3LO}.

The largest contributions to the neutron-matter energy arise from NN
interactions. In Refs.~\cite{N3LO,longN3LO} we studied the
perturbative convergence of all existing NN potentials at N$^2$LO and
at N$^3$LO of Epelbaum, Gl\"ockle, and Mei{\ss}ner
(EGM)~\cite{EGM,EGM2} with cutoffs $\Lambda/ \tilde{\Lambda} =
450/500$, $450/700$, $550/600$, $600/600$, and $600/700 \mev$, where
$\Lambda$ and $\tilde{\Lambda}$ denote the cutoff in the
Lippmann-Schwinger equation and in the two-pion-exchange
spectral-function regularization, respectively; as well as the
available N$^3$LO NN potentials of Entem and Machleidt
(EM)~\cite{EMRept,EM} with cutoffs $\Lambda= 500$ and $600 \mev$.

\begin{figure}[t]
\centerline{\includegraphics[width=1.0\textwidth,clip=]{%
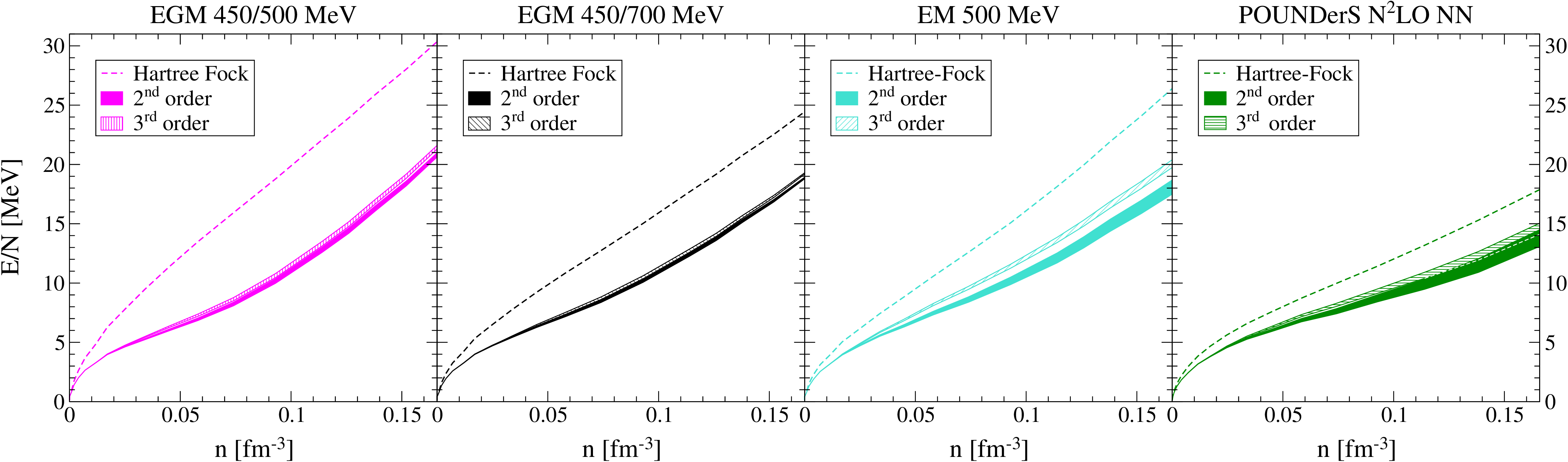}}
\caption{Panel 1-3:~Neutron-matter energy per particle $E/N$
as a function of density $n$ for the N$^3$LO NN potentials that
exhibit a perturbative convergence. The dashed lines are
Hartree-Fock results. The filled and shaded bands are second- and
third-order results, where at each order the band
ranges from using a free to a Hartree-Fock spectrum. All calculations
include N$^2$LO 3N forces with a 3N cutoff $\Lambda=2.0 \fmi$ and
low-energy couplings $c_1 = 0.75 \gevi$ and $c_3 = 4.77\gevi$. For
details see Ref.~\cite{longN3LO}. Panel 4:~Same for the POUNDerS
N$^2$LO NN potential (without 3N forces).\label{fig:convergence}}      
\end{figure}

To study the perturbative convergence of the different NN potentials,
we calculated the Hartree-Fock as well as second- and third-order
energies, only including particle-particle diagrams, with both free
and Hartree-Fock single-particle energies. The results for NN and
N$^2$LO 3N forces are shown in Fig.~\ref{fig:convergence} for the
perturbative NN interactions for a 3N cutoff $\Lambda = 2.0 \fmi$ and
a particular choice of $c_1 = 0.75 \gevi$, $c_3 = 4.77 \gevi$,
although the general picture is unchanged for other coupling
values. The bands result from using a free to a Hartree-Fock
single-particle spectrum. The N$^3$LO EGM potentials with cutoffs
$450/500 \mev$ and $450/700 \mev$ exhibit only small energy changes
from second to third order. This indicates that these potentials are
perturbative for neutron matter. For the EM $500 \mev$ potential the
difference between second and third order is larger compared to the
EGM potentials. Since this potential is most commonly used in nuclear
structure calculations, we include it in our complete N$^3$LO
calculation. The perturbative convergence for these potentials in
neutron matter is similar to renormalization-group-evolved
interactions in nuclear matter~\cite{nucmatt}. We have also studied in
Fig.~\ref{fig:convergence} the POUNDerS N$^2$LO NN
potential~\cite{POUNDerS}, which is found to be perturbative as well.
In addition, there are in-medium chiral perturbation theory schemes
that treat the Fermi momentum as an explicit
scale~\cite{Weise,Lacour}.

The larger-cutoff N$^3$LO EGM $550/600 \mev$ and $600/600 \mev$
potentials as well as the EM $600 \mev$ potential are not used in our
calculations because they show large changes from second to third
order~\cite{longN3LO}. This demonstrates that these interactions are
nonperturbative. The N$^3$LO EGM $600/600 \mev$ potential is not used
because it breaks Wigner symmetry ($C_T=0$) at the interaction level
(as discussed in Ref.~\cite{longN3LO}).

The subleading N$^3$LO 3N forces have been derived
recently~\cite{N3LOlong,N3LOshort}.  They can be grouped into five
topologies, where the latter two depend on the NN contacts $C_{T/S}$:
\begin{equation}
V_{\text{3N}}^{\text{N}^3\text{LO}} = V^{2\pi} + V^{2\pi\text{-}1\pi} + 
V^{\text{ring}} + V^{2\pi\text{-cont}} + V^{1/m} \,.
\end{equation}
$V^{2\pi}$, $V^{2\pi\text{-}1\pi}$, and $V^{\text{ring}}$ denote the
long-range two-pion-exchange, the two-pion--one-pion-exchange, and the
pion-ring 3N interactions, respectively~\cite{N3LOlong}. The terms
$V^{2\pi\text{-cont}}$ and $V^{1/m}$ are the short-range
two-pion-exchange--contact 3N interaction and 3N relativistic
corrections~\cite{N3LOshort}. The N$^3$LO 4N forces have been derived
in Refs.~\cite{4N,4Nlong} and in general depend on the contact $C_T$,
but in neutron matter the $C_T$-dependent parts do not
contribute. There are seven 4N topologies that lead to non-vanishing
contributions. In neutron matter only two three-pion-exchange diagrams
(in Ref.~\cite{4N} named $V^a$ and $V^e$) and the
pion-pion-interaction diagram ($V^f$) contribute~\cite{N3LO}.

\begin{figure}[t]
\centerline{\includegraphics[width=\textwidth,clip=]{%
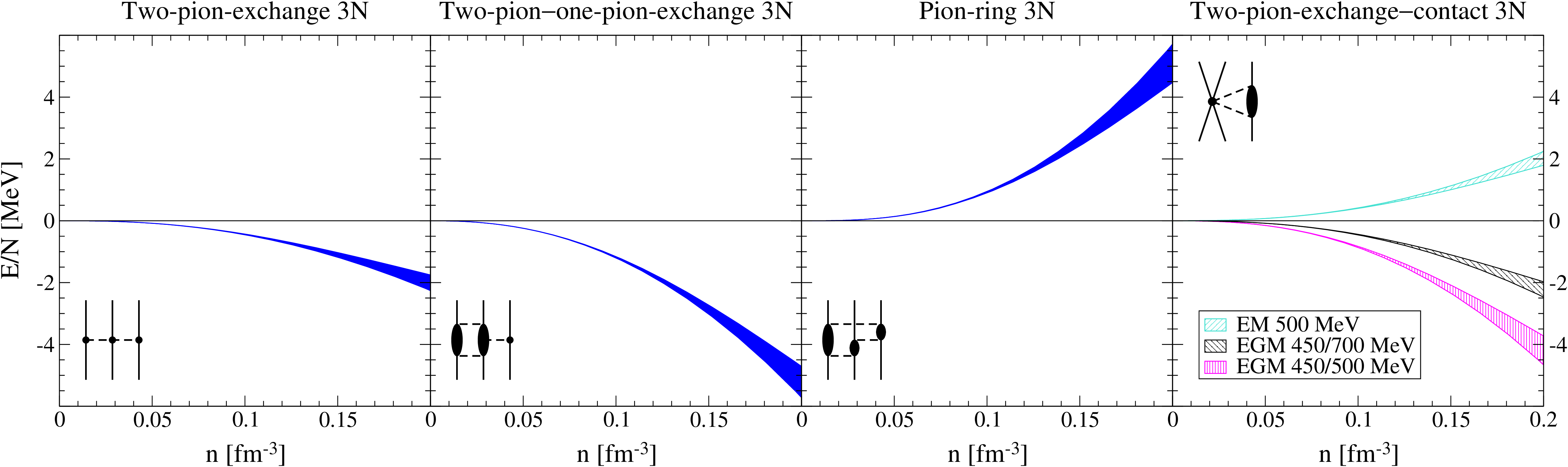}}
\vspace*{0.2cm}
\centerline{\includegraphics[width=\textwidth,clip=]{%
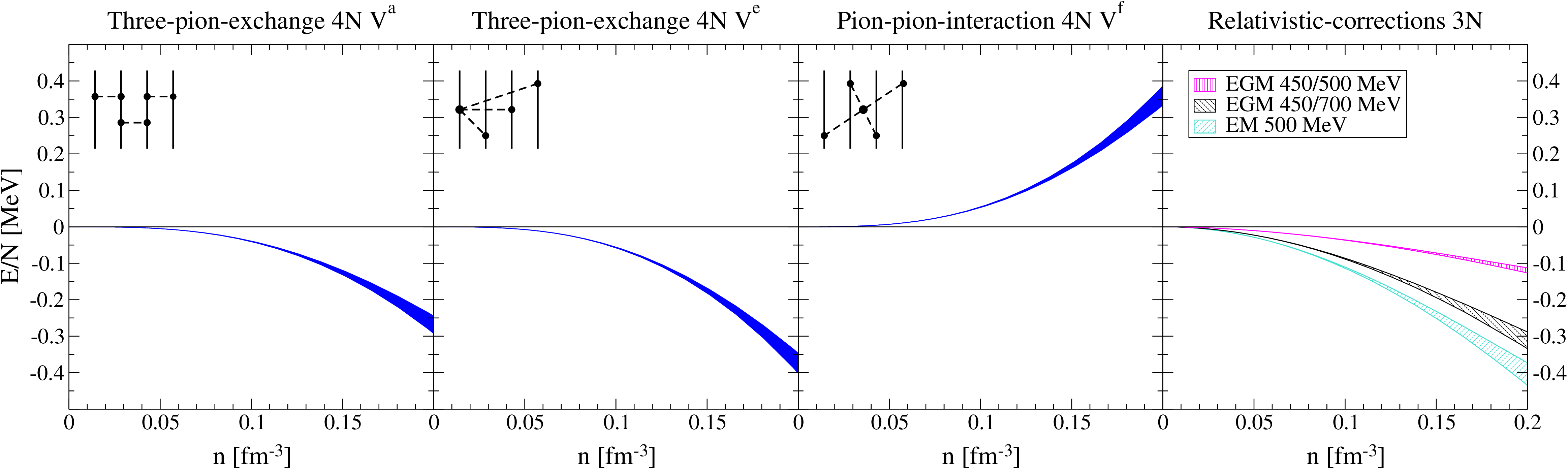}}
\caption{Energy per particle $E/N$ as a function of density $n$
for all individual N$^3$LO 3N- and 4N-force contributions to neutron
matter at the Hartree-Fock level~\cite{N3LO}. All bands are obtained
by varying the 3N/4N cutoffs $\Lambda = 2.0 - 2.5 \fmi$. For the
3N two-pion-exchange--contact forces and the 3N relativistic corrections,
the different bands correspond to the different NN contacts, $C_T$ and
$C_S$, determined consistently for the N$^3$LO EM/EGM potentials. The
diagrams illustrate the 3N/4N force topology.\label{fig:individualnm}}
\end{figure}

The N$^3$LO many-body interactions are evaluated in the Hartree-Fock
approximation, which is expected to be reliable for neutron
matter~\cite{nm}. We show the individual contributions of the 3N and
4N forces in Fig.~\ref{fig:individualnm}, where the bands correspond
to the 3N/4N cutoff variation $\Lambda =2-2.5 \fmi$. The N$^3$LO
two-pion-exchange expectation value (panel 1) sets the expected scale
of N$^3$LO 3N interactions. Compared to this, we find relatively large
expectation values in the $V^{2\pi\text{-}1\pi}$, $V^{\text{ring}}$,
and $V^{2\pi\text{-cont}}$ topologies. This could indicate that in
these topologies $\Delta$ contributions shifted to N$^4$LO are
expected to be important~\cite{longN3LO,Krebs}. The 3N relativistic
corrections and the contributions from N$^3$LO 4N forces are small
(see also Ref.~\cite{Fiorilla}). However, also for 4N forces
additional larger contributions from $\Delta$ excitations may arise at
N$^4$LO~\cite{Kaiser}.

\begin{figure}[t]
\centerline{\includegraphics[width=0.48\textwidth,clip=]{%
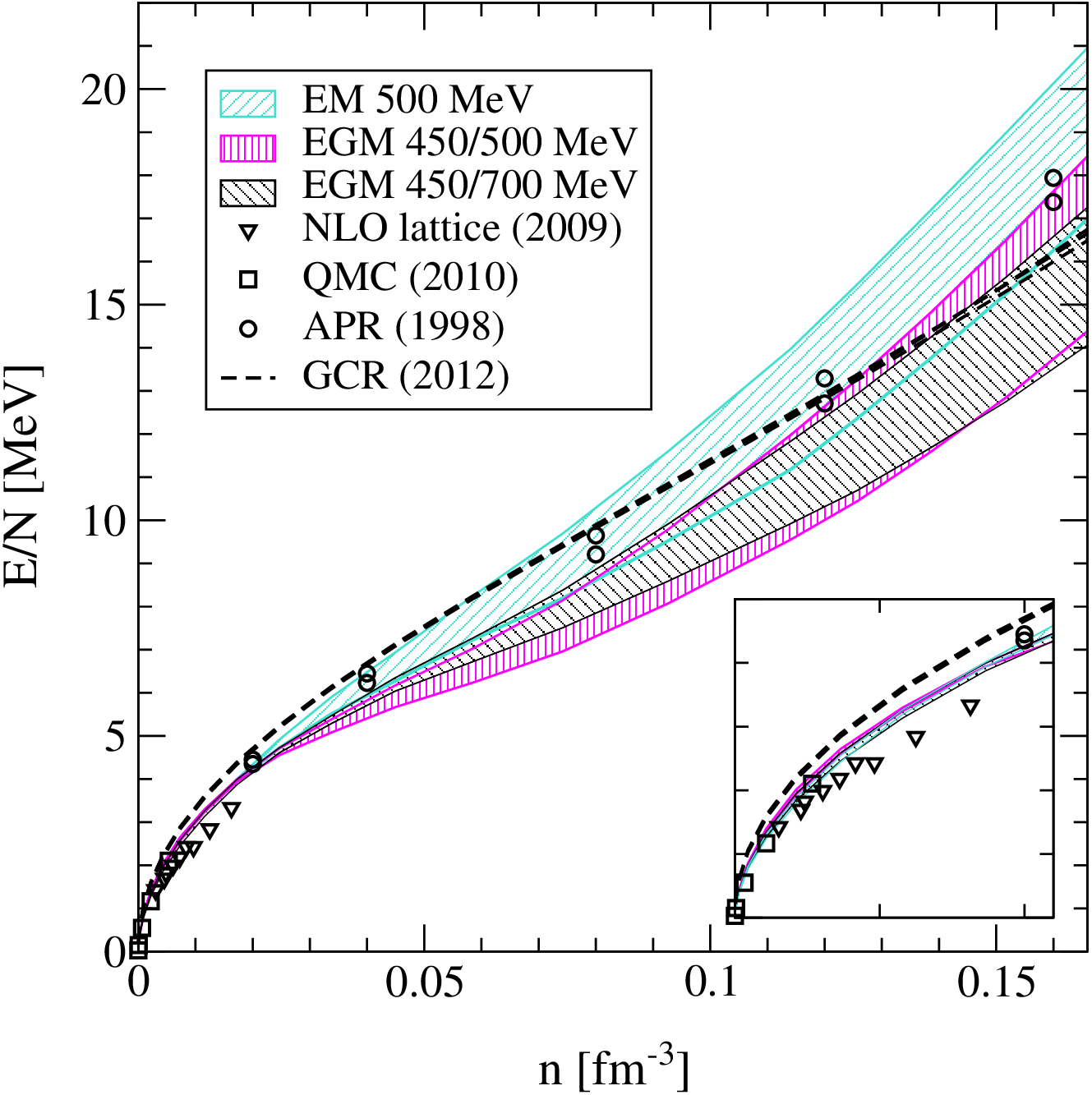}
\hspace*{0.2cm}
\includegraphics[width=0.48\textwidth,clip=]{%
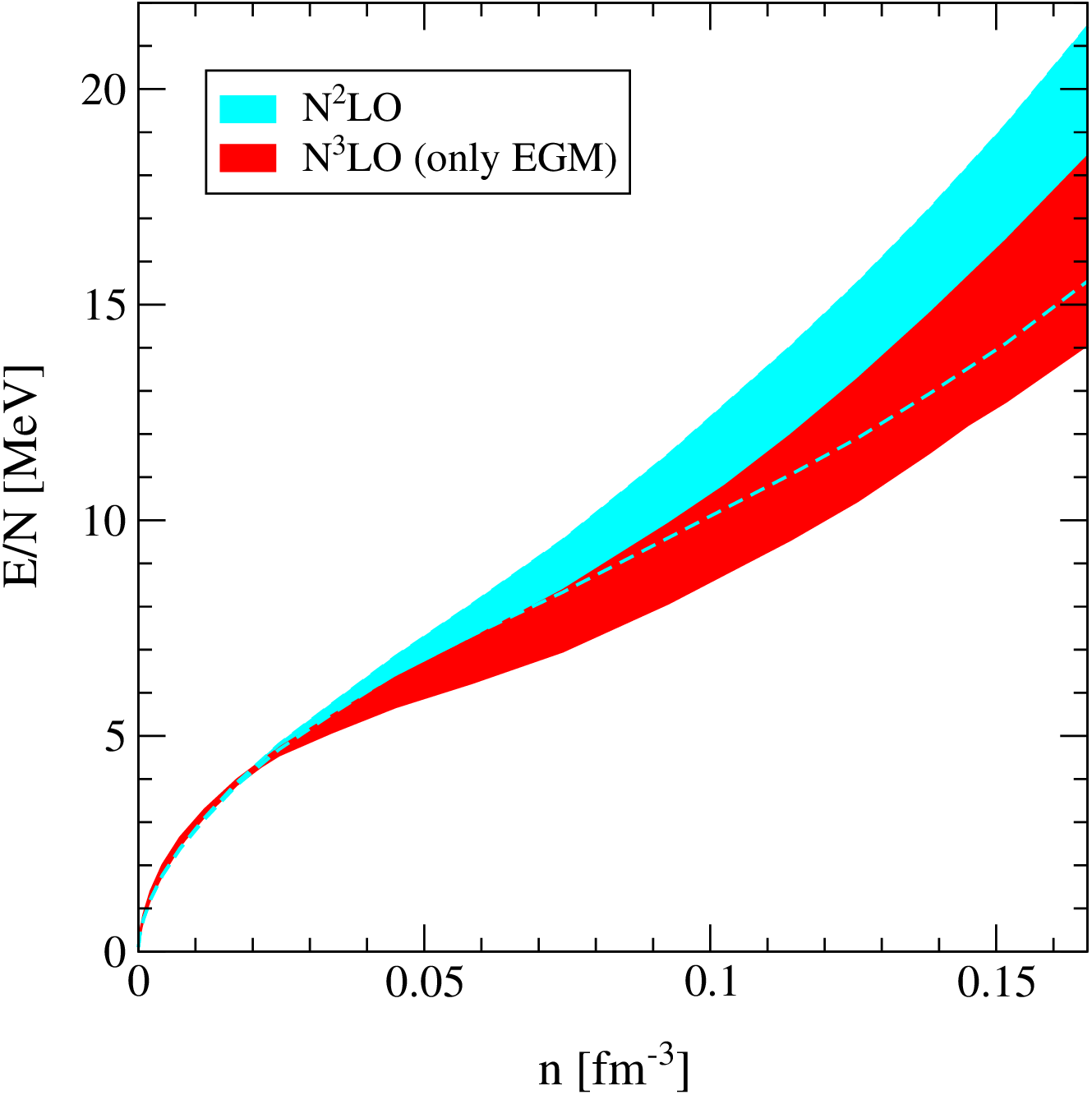}}
\caption{Left panel:~Neutron-matter energy per particle $E/N$ as a
function of density $n$ including NN, 3N and 4N forces to N$^3$LO for
the given EM/EGM NN potentials~\cite{N3LO}. The bands include
uncertainty estimates due to the many-body calculation, the low-energy
$c_i$ couplings, and by varying the 3N/4N cutoffs. For comparison,
results are shown at low densities (see also the inset) from NLO
lattice~\cite{NLOlattice} and QMC simulations~\cite{GC}, and at
nuclear densities from variational~\cite{APR} and Auxiliary Field
Diffusion MC calculations (GCR)~\cite{GCR} based on phenomenological
potentials. Right panel:~Neutron-matter energy per particle at N$^2$LO
(upper blue band that extends to the dashed line) and N$^3$LO (lower
red band)~\cite{N3LO}. The bands are based on the EGM NN potentials
and include the same uncertainty estimates.\label{fig:N3LO}}
\end{figure}

The complete N$^3$LO result for neutron matter is shown in the left
panel of Fig.~\ref{fig:N3LO}, which includes all NN, 3N, and 4N
interactions to N$^3$LO~\cite{N3LO}. At saturation density, we obtain
for the energy per particle $E/N = 14.1-21.0 \mev$. This range is
based on different NN potentials, a variation of the couplings $c_1 =
-(0.75 - 1.13)\gevi$, $c_3 = -(4.77 - 5.51) \gevi$ \cite{Krebs}, which
dominates the total uncertainty, a 3N/4N-cutoff variation $\Lambda = 2
- 2.5 \fmi$, and the uncertainty in the many-body calculation.

The neutron-matter energy in Fig.~\ref{fig:N3LO} is in very good
agreement with NLO lattice results~\cite{NLOlattice} and QMC
simulations~\cite{GC} at very low densities (see also the inset). At
nuclear densities, we compare our N$^3$LO results with variational
calculations based on phenomenological potentials (APR)~\cite{APR},
which are within the N$^3$LO band, but do not provide theoretical
uncertainties. In addition, we compare the density dependence with
results from Auxiliary Field Diffusion MC (AFDMC) calculations
(GCR)~\cite{GCR} based on nuclear force models adjusted to a
symmetry energy of $32 \mev$.

We also compare the convergence from N$^2$LO to N$^3$LO in the same
calculational setup. For this comparison, we only consider the EGM
potentials with cutoffs $450/500 \mev$ and $450/700 \mev$. This leads
to an N$^3$LO energy range of $14.1 - 18.4 \mev$ per particle at
$n_0$. For the N$^2$LO band in the right panel of Fig.~\ref{fig:N3LO},
we have estimated the theoretical uncertainties in the same way, and
find an energy of $15.5 - 21.4 \mev$ per particle at $n_0$. The two
bands overlap but the range of the band is only reduced by a factor of
$2/3$ in contrast to the $1/3$ expected from the EFT power
counting. We attribute this to $\Delta$ effects (see the discussion in
Refs.~\cite{N3LO,longN3LO}).

\section{QMC calculations with chiral EFT interactions}

Quantum Monte Carlo methods have not been used with chiral EFT
interactions due to nonlocalities in their present implementation in
momentum space. Nonlocalities are difficult to handle in
QMC~\cite{QMC}. In the momentum-space interactions, there are two
sources of nonlocalities: first, due to regulator functions that lead
to nonlocal interactions upon Fourier transformation, and second, due
to contact interactions that depend on the momentum transfer in the
exchange channel ${\bf k}$ and from ${\bf k}$-dependent parts in
pion-exchange contributions beyond N$^2$LO. For applications in QMC,
we have developed local chiral EFT interactions in Ref.~\cite{QMC}.

To avoid regulator-generated nonlocalities for the long-range
pion-exchange parts, we use the local coordinate-space expressions for
the LO one-pion-exchange as well as NLO and N$^2$LO two-pion-exchange
interactions~\cite{DR,SF} and regulate them directly in coordinate
space using the function $f_{\rm long}(r) = 1 - e^{-(r/R_0)^4}$, which
smoothly cuts off interactions at short distances $r < R_0$ while
leaving the long-range parts unchanged~\cite{QMC}. So, $R_0$ takes
over the role of the cutoff $\Lambda$ in momentum space.

To remove the ${\bf k}$-dependent contact interactions to N$^2$LO, we
make use of the freedom to choose a basis of short-range operators in
chiral EFT interactions (similar to Fierz ambiguities). At LO, one
usually considers the two momentum-independent contact interactions
$C_S + C_T \, {\bm \sigma}_1 \cdot {\bm \sigma}_2$. However, it is
equivalent to choose any two of the four operators $\mathbbm{1}, \, {\bm
\sigma}_1 \cdot {\bm \sigma}_2, \, {\bm \tau}_1 \cdot {\bm \tau}_2$,
and ${\bm \sigma}_1 \cdot {\bm \sigma}_2 \, {\bm \tau}_1 \cdot {\bm
\tau}_2$, with spin and isospin operators ${\bm \sigma}_i, {\bm
\tau}_i$, because there are only two S-wave channels due to the
Pauli principle. It is a convention in present chiral EFT interactions
to neglect the isospin dependence, which is then generated from the
exchange terms~\cite{EGM,EGM2,EM}.

We use this freedom to keep at NLO (order $Q^2$) an isospin-dependent
$q^2$ contact interaction and an isospin-dependent $({\bm \sigma}_1
\cdot {\bf q})({\bm \sigma}_2 \cdot {\bf q})$ tensor part in favor of
a nonlocal $k^2$ contact interaction and a nonlocal $({\bm \sigma}_1
\cdot {\bf k})({\bm \sigma}_2 \cdot {\bf k})$ tensor part. This leads
to the following seven linearly independent contact interactions at
NLO that are local~\cite{QMC},
\begin{align}
V^{\rm NLO}_{\rm short} &= C_1 \, q^2 + C_2 \, q^2 \, 
{\bm \tau}_1 \cdot {\bm \tau}_2 \nonumber 
+ \bigl(C_3 \, q^2 + C_4 \, q^2 \, {\bm \tau}_1 \cdot {\bm \tau}_2 \bigr)
\, {\bm \sigma}_1 \cdot {\bm \sigma}_2 \nonumber \\
&+ i \, \frac{C_5}{2} \, ({\bm \sigma}_1 + {\bm \sigma}_2) \cdot
{\bf q} \times {\bf k} \nonumber 
+ C_6 \, ({\bm \sigma}_1 \cdot {\bf q})({\bm \sigma}_2 \cdot {\bf q}) 
\nonumber + C_7 \, ({\bm \sigma}_1 \cdot {\bf q})({\bm \sigma}_2 \cdot {\bf q}) 
\, {\bm \tau}_1 \cdot {\bm \tau}_2 \,,
\label{eq:NLOshort}
\end{align}
where the only ${\bf k}$-dependent contact interaction ($C_5$) is a
spin-orbit potential.

\begin{figure}[t]
\centerline{\includegraphics[width=\textwidth,clip=]{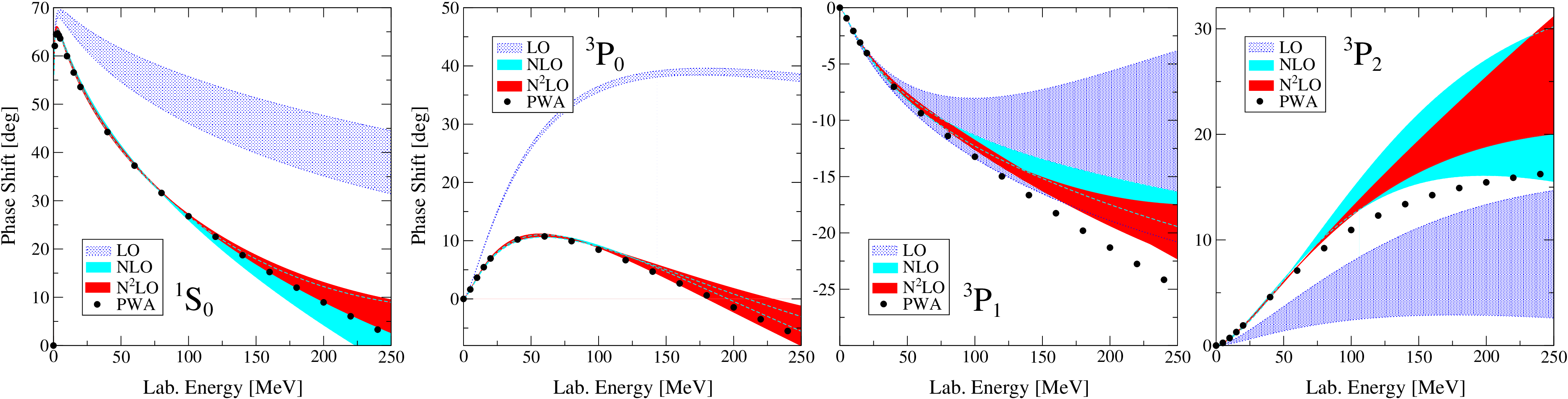}}
\caption{Neutron-proton phase shifts as a function of laboratory 
energy in the $^1$S$_0$, $^3$P$_0$, $^3$P$_1$, and $^3$P$_2$ partial
waves in comparison to the Nijmegen partial-wave analysis
(PWA)~\cite{Stoks:1993}. The LO, NLO, and N$^2$LO local chiral
potential bands are obtained by varying $R_0$ between $0.8-1.2 \, 
{\rm fm}$ (with a spectral-function cutoff $\widetilde{\Lambda}=800 \,
{\rm MeV}$)~\cite{QMC,Freunek:2007}.\label{fig:phases}}
\end{figure}

\begin{figure}[t]
\centerline{\includegraphics[width=0.48\textwidth,clip=]{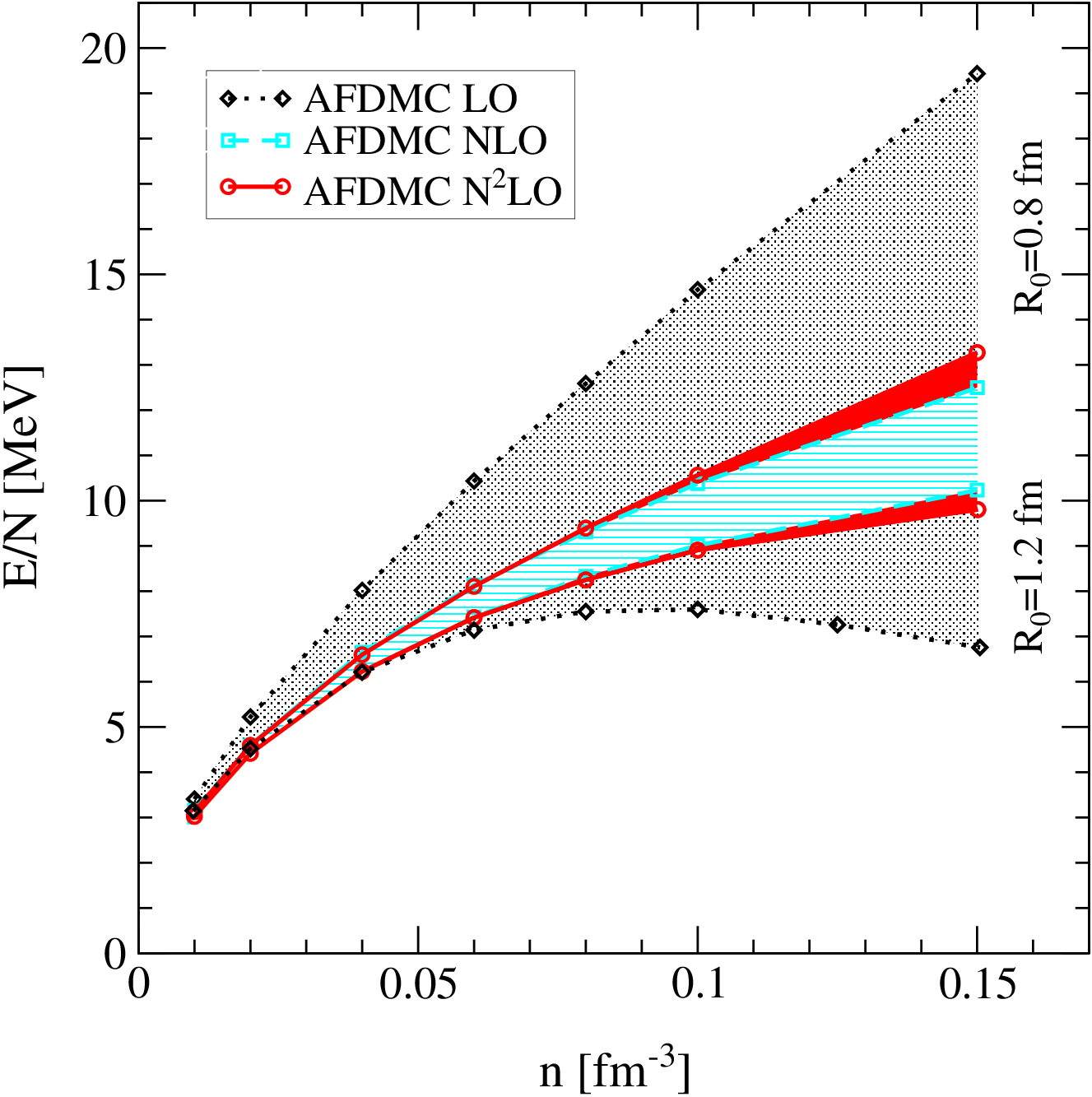}
\hspace*{0.2cm}
\includegraphics[width=0.48\textwidth,clip=]{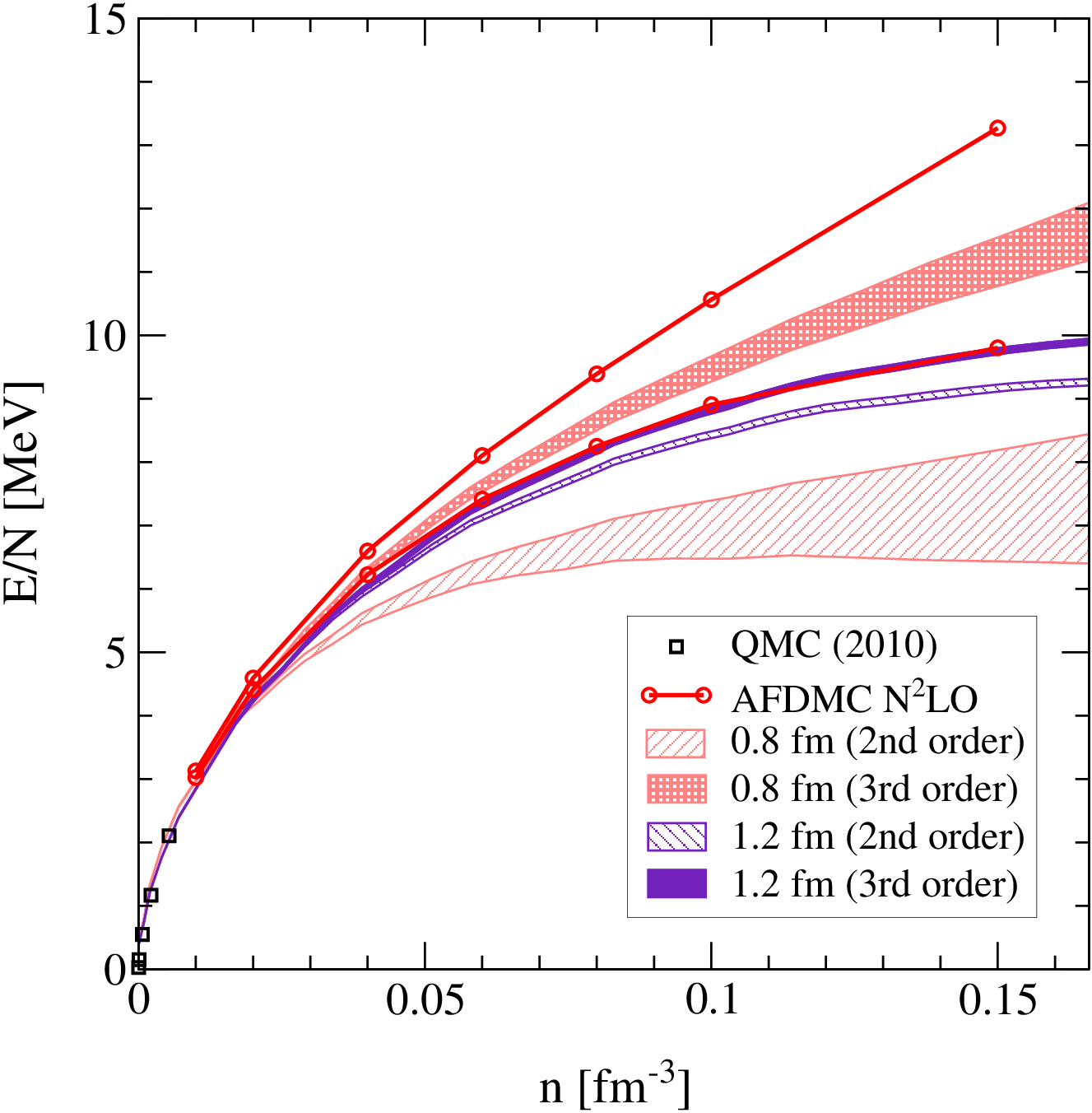}}
\caption{Left panel:~Neutron-matter energy per particle $E/N$ as a
function of density $n$ calculated using AFDMC with chiral EFT NN
interactions at LO, NLO, and N$^2$LO~\cite{QMC}. The statistical errors are
smaller than the points shown. The lines give the range obtained by
varying $R_0$ between $0.8-1.2 \, {\rm fm}$. Right panel:~The AFDMC
N$^2$LO band in comparison to perturbative calculations using the same
N$^2$LO NN interactions. The lower (upper) limit of the AFDMC
N$^2$LO band is for $R_0 = 1.2 \, {\rm fm}$ ($R_0 = 0.8 \, {\rm fm}$),
corresponding to a momentum cutoff $\Lambda \sim 400 \, {\rm MeV}$
($\Lambda \sim 600 \, {\rm MeV}$). Perturbative results are shown at
second and third order. For the softer $R_0 = 1.2 \, {\rm fm}$
interaction (narrow purple bands), third-order corrections are
small and the third-order energy is in excellent agreement with the
AFDMC results, while for the harder $R_0 = 0.8 \, {\rm fm}$
interaction (light red bands), the convergence is clearly
slow.\label{fig:eos_qmc}}
\end{figure}

The low-energy couplings $C_{S/T}$ at LO plus $C_{1-7}$ at NLO and
N$^2$LO are fit in Ref.~\cite{Freunek:2007} for different $R_0$ to the
NN phase shifts of the Nijmegen partial-wave
analysis~\cite{Stoks:1993} at laboratory energies $E_{\rm lab} = 1, 5,
10, 25, 50,$ and $100 \, {\rm MeV}$, using a local regulator. The
reproduction of the isospin $T=1$ S- and P-waves is shown order by
order in Fig.~\ref{fig:phases}, where the bands are obtained by
varying $R_0$ between $0.8-1.2 \, {\rm fm}$ and provide a measure of
the theoretical uncertainty. At N$^2$LO, an isospin-symmetry-breaking
contact interaction ($C_{nn}$ for neutrons) is added, which is fit to
$a_{nn} = -18.8 \, {\rm fm}$. As shown in Fig.~\ref{fig:phases}, the
comparison with NN phase shifts is very good for $E_{\rm lab} \lesssim
150 \, {\rm MeV}$. This is similar for higher partial waves and
isospin $T=0$ channels. In cases where there are deviations for higher
energies (such as in the $^3$P$_2$), the width of the band signals
significant theoretical uncertainties due to the chiral EFT truncation
at N$^2$LO. The NLO and N$^2$LO bands nicely overlap or are very
close, but it is also apparent that the bands at N$^2$LO are of a
similar size as at NLO. This is because the width of the bands at both
NLO and N$^2$LO shows effects of the neglected order-$Q^4$ contact
interactions.

Since nuclear forces contain quadratic spin, isospin, and tensor
operators (of the form ${\bm \sigma}^{\alpha}_i \, A^{\alpha \beta}_{ij}
\, {\bm \sigma}^{\beta}_j$), the many-body wave function cannot be
expressed as a product of single-particle spin-isospin states. All
possible spin-isospin nucleon-pair states need to be explicitly
accounted for, leading to an exponential increase in the number of
possible states. However, the AFDMC method~\cite{Schmidt:1999} is
capable of efficiently handling spin-dependent Hamiltonians. AFDMC
rewrites the Green's function by applying a Hubbard-Stratonovich
transformation using auxiliary fields to change the quadratic
spin-isospin operator dependences to linear. For the case of neutrons,
it is also possible to include spin-orbit interactions and 3N
forces in AFDMC nonperturbatively~\cite{Sarsa:2003,Gandolfi:2009}.

In the left panel of Fig.~\ref{fig:eos_qmc} we show first AFDMC
calculations for the neutron-matter energy with local chiral EFT NN
interactions at LO, NLO, and N$^2$LO~\cite{QMC}. At each order, the
full interaction is used both in the propagator and when evaluating
observables.  The bands in Fig.~\ref{fig:eos_qmc} give the range of
the energy obtained by varying $R_0$ between $0.8-1.2 \, {\rm fm}$,
where the softer $R_0 = 1.2 \, {\rm fm}$ interactions yield the lower
energies.  At LO, the energy has a large uncertainty. The overlap of
the bands at different orders in Fig.~\ref{fig:eos_qmc} is very
systematic. In addition, the result that the NLO and N$^2$LO bands are
comparable is expected from the discussion of the phase-shift bands in
Fig.~\ref{fig:phases} and from the large $c_i$ entering at N$^2$LO.

Our AFDMC results provide first nonperturbative benchmarks for chiral
EFT interactions at nuclear densities. We have performed perturbative
calculations as in the previous section based on the same local
N$^2$LO NN interactions. The perturbative energies are compared in the
right panel of Fig.~\ref{fig:eos_qmc} to the AFDMC N$^2$LO
results. For the softer $R_0 = 1.2 \, {\rm fm}$ ($\Lambda \sim 400 \,
{\rm MeV}$) interaction, the third-order corrections are small and the
perturbative third-order energy is in excellent agreement with the
AFDMC results, while for the harder $R_0 = 0.8 \, {\rm fm}$ ($\Lambda
\sim 600 \, {\rm MeV}$) interaction, the convergence is clearly
slow. This is the first nonperturbative validation for neutron matter
of the possible perturbativeness of low-cutoff $\Lambda \sim 400 \,
{\rm MeV}$ interactions~\cite{PPNP}.

\section{Astrophysical applications}

\begin{figure}[t]
\centerline{\includegraphics[width=0.48\textwidth,clip=]{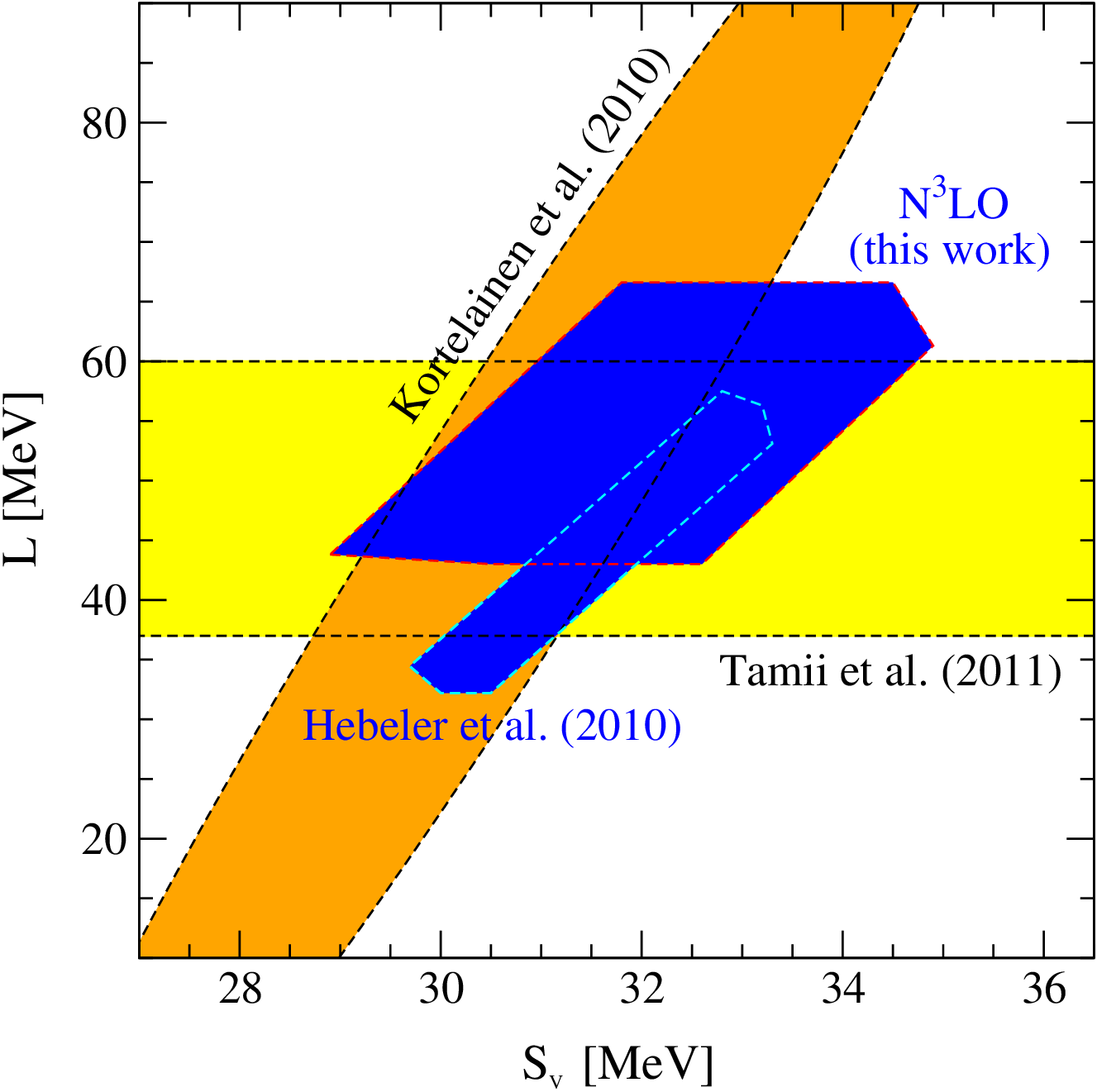}
\hspace*{0.2cm}
\includegraphics[width=0.48\textwidth,clip=]{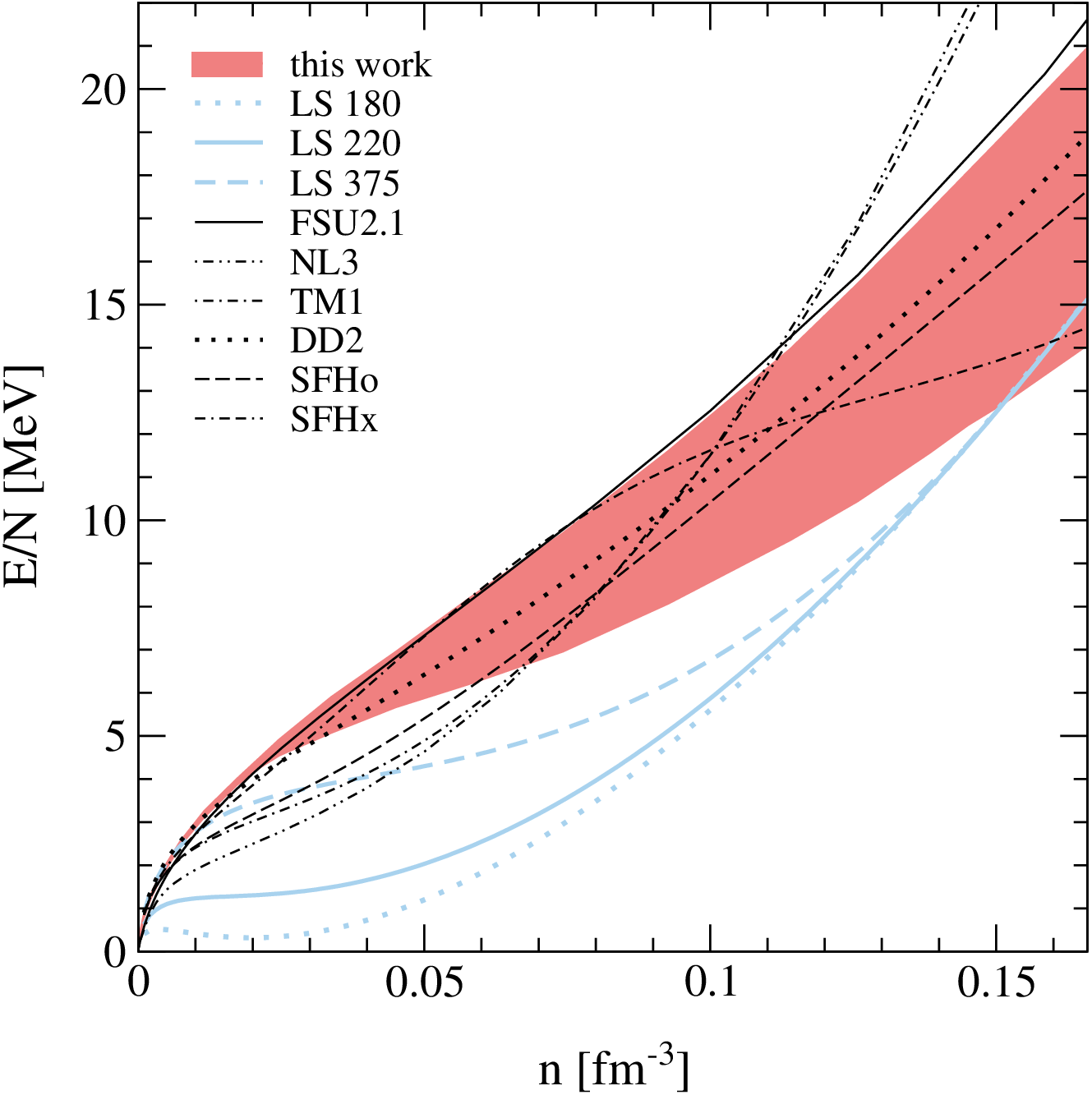}}
\caption{Left panel:~Range for the symmetry energy $S_v$ and its
density dependence~$L$ obtained at N$^3$LO~\cite{N3LO} versus
including 3N forces at N$^2$LO (Hebeler {\it et al.}~\cite{nstar_long}).
For comparison, see Ref.~\cite{LL}, we show constraints obtained from
energy-density functionals for nuclear masses (Kortelainen {\it
et al.}~\cite{Kortelainen2010}) and from the $^{208}$Pb dipole
polarizability (Tamii {\it et al.}~\cite{Tamii2011}). Right
panel:~Comparison of the N$^3$LO neutron-matter energy of the left
panel of Fig.~\ref{fig:N3LO} (red band) with equations of state for
core-collapse supernova simulations provided by Lattimer-Swesty (LS
with different incompressibilities 180, 220, and $375 \, {\rm MeV}$),
G.~Shen (FSU2.1, NL3), Hempel (TM1, SFHo, SFHx), and Typel (DD2). For
details see Ref.~\cite{longN3LO}.\label{SvL}}
\end{figure}

The symmetry energy $S_v$ and its density derivative $L$ provide
important input for astrophysics~\cite{LL}. To calculate these, we
follow Ref.~\cite{nstar_long}. The predicted ranges for $S_v$ and $L$
at saturation density are $S_v=28.9 - 34.9 \, {\rm MeV}$ and $L=43.0 -
66.6 \, {\rm MeV}$. The $S_v$ and $L$ ranges are in very good
agreement with experimental constraints from nuclear
masses~\cite{Kortelainen2010} and from the dipole polarizability of
$^{208}$Pb~\cite{Tamii2011}, see the left panel of Fig.~\ref{SvL}. In
addition, they also overlap with the results for RG-evolved NN
interactions with N$^2$LO 3N forces~\cite{LL,nstar_long}, but due to
the additional density dependences from N$^3$LO many-body forces, the
correlation between $S_v$ and $L$ is not as tight.

The neutron-matter results also provide constraints for equations of
state for core-collapse-supernova simulations. In the right panel of
Fig.~\ref{SvL}, we compare the N$^3$LO neutron-matter band (red band)
to the Lattimer-Swesty (LS) equation of state~\cite{Lattimer1991}
(with different incompressibilities 180, 220, and $375$\,MeV), which
is most commonly used in simulations, and to different relativistic
mean-field-theory equations of state based on the density functionals
DD2~\cite{Typelprivate}, FSU2.1~\cite{GShen2011oc},
NL3~\cite{GShen2011t}, SFHo, SFHx~\cite{Steiner2012}, and
TM1~\cite{HShen2011}. At low densities only the DD2, FSU2.1 and SFHx
equations of state are consistent with the N$^3$LO neutron-matter
band. The NL3 and TM1 equations of state have a too strong density
dependence, which leads to unnaturally large $S_v$ and $L$ values. In
addition, Fig.~\ref{SvL} exhibits a strange density dependence of SFHx.

\begin{figure}[t]
\centerline{\includegraphics[width=0.48\textwidth,clip=]{%
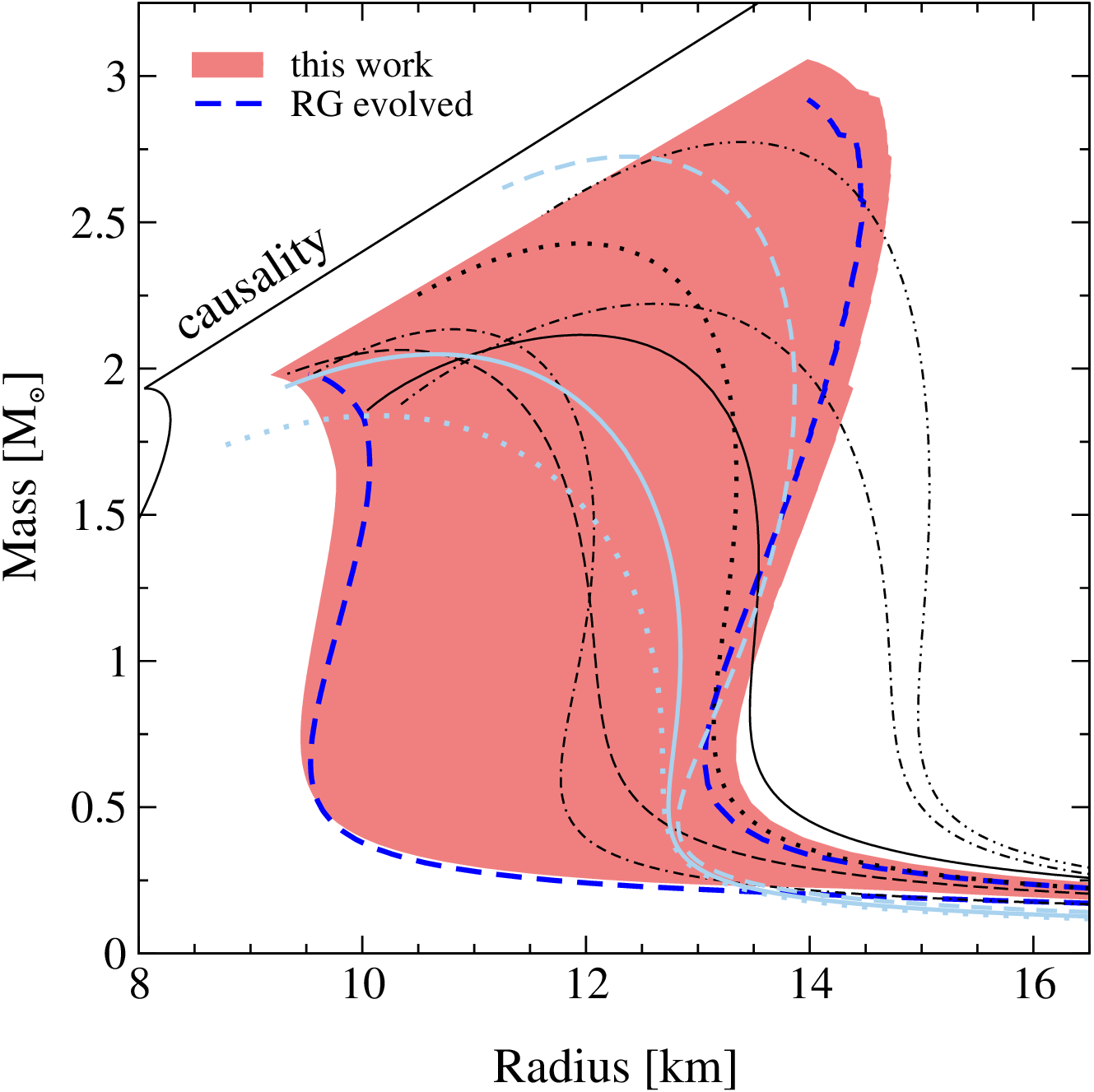}
\hspace*{0.2cm}
\includegraphics[width=0.48\textwidth,clip=]{%
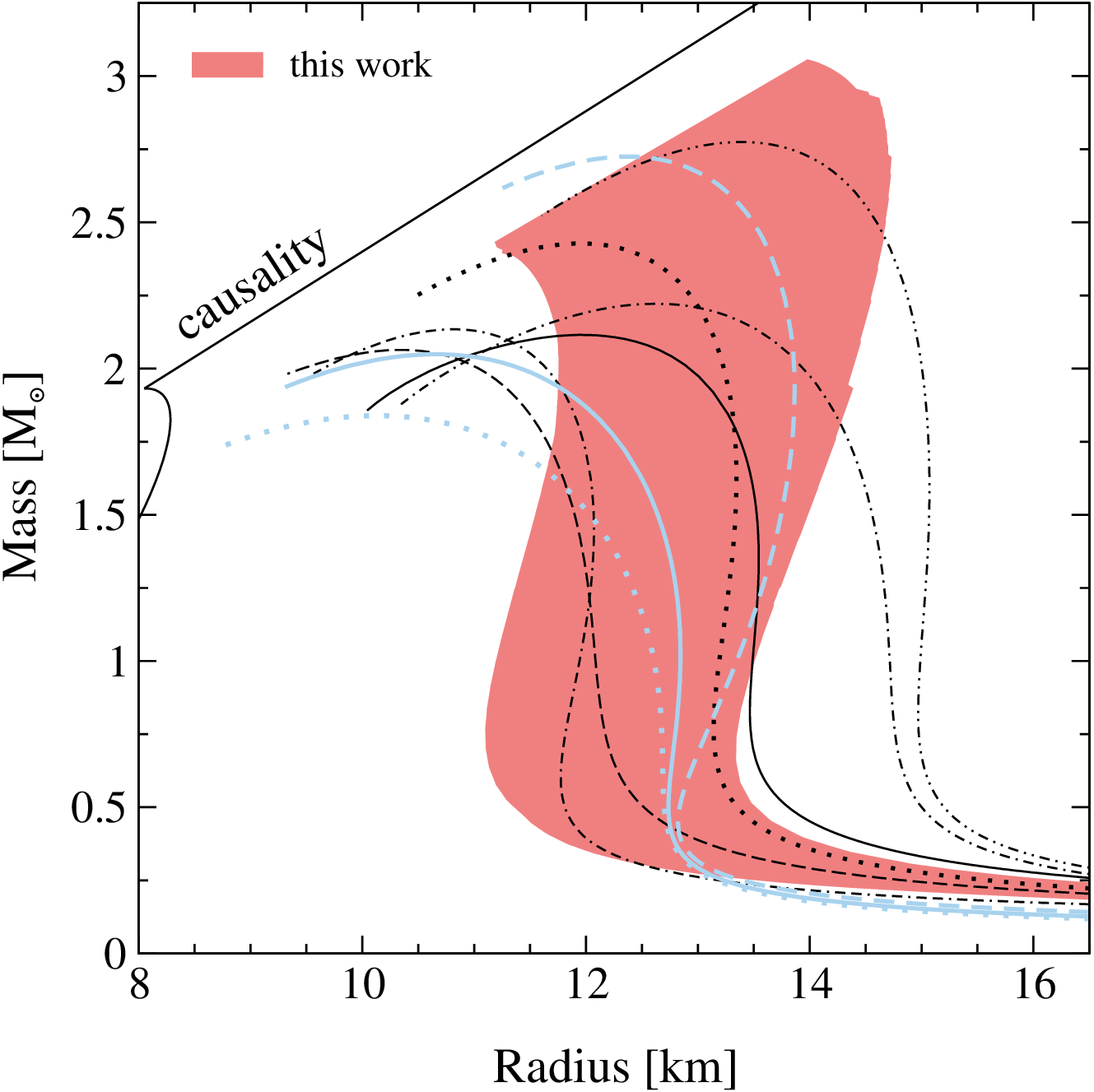}}
\caption{Constraints on the mass-radius diagram of neutron stars based
on our neutron-matter results at N$^3$LO following
Ref.~\cite{nstar_long,nstar} for the extension to neutron-star matter
and to high densities (red band), in comparison to the constraints
from calculations based on RG-evolved NN interactions (thick dashed
blue lines)~\cite{nstar_long}.  We also show the mass-radius relations
obtained from the equations of state for core-collapse supernova
simulations shown in Fig.~\ref{SvL}.  Left panel:~Band obtained with
the constraint of a $1.97\, M_{\odot}$ neutron star~\cite{longN3LO}.
Right panel:~Same for a $2.4 \, M_{\odot}$ star.\label{fig:nstar}}
\end{figure}

Next, we use the N$^3$LO neutron-matter results to provide constraints
for the structure of neutron stars. We follow
Refs.~\cite{nstar_long,nstar} for incorporating beta equilibrium and
for the extension to high densities using piecewise polytropes that
are constrained by causality and by the requirement to support a $1.97
\pm 0.04 \, M_\odot$ neutron star~\cite{Demorest} (see also the recent
$2.01 \pm 0.04 \, M_\odot$ discovery~\cite{Antoniadis}). In addition,
we consider the case, if a $2.4 \, M_\odot$ neutron star were to be
observed. The resulting constraints on the neutron star mass-radius
diagram are shown in Fig.~\ref{fig:nstar} by the red bands. The bands
represent an envelope of a large number of individual equations of
state reflecting the uncertainties in the N$^3$LO neutron-matter
calculation and in the polytropic extensions to high
densities~\cite{nstar_long,nstar}. The combination with the $2 \,
M_\odot$ neutron star (left panel) predicts a radius range of
$9.7-13.9$ km for a $1.4 \, M_\odot$ star~\cite{longN3LO,nstar_long}.
The maximal neutron star mass is found to be $3.1 \, M_\odot$, with a
corresponding radius of about $14$ km. We also find very good
agreement with the mass-radius constraints from the neutron-matter
calculations based on RG-evolved NN interactions with N$^2$LO 3N
forces~\cite{nstar_long}, which are shown by the thick dashed blue
lines in the left panel of Fig.~\ref{fig:nstar}.

In addition, we show in Fig.~\ref{fig:nstar} the mass-radius relations
obtained from equations of state for core-collapse supernova
simulations~\cite{Lattimer1991,GShen2011oc,GShen2011t,%
Steiner2012,HShen2011,Hempel2012,ConnorKleiner}. The inconsistency
in Fig.~\ref{SvL} of many of the equations of state with the N$^3$LO
neutron-matter band at low densities results in a large spread of very
low mass/large radius neutron stars, where the red band is
considerably narrower in Fig.~\ref{fig:nstar}. For typical neutron
stars, our calculations rule out the NL3 and TM1 equations of state,
which produce too large radii. Finally, we have also explored the
constraints from N$^3$LO calculations for the chiral condensate in
neutron matter~\cite{chiralcond}.

\vspace*{0.5cm}
\noindent
All the very best for your 70th birthday, James, lots of good health
and energy for fun in life and physics (and many days like the one we
enjoyed in Capri)! We would like to thank E.~Epelbaum, S.~Gandolfi,
J.~M.~Lattimer, A.~Nogga, and C.~J.~Pethick, who contributed to the
results presented in this talk. This work was supported by the DFG
through Grant SFB 634, the ERC Grant No.~307986 STRONGINT, the
Helmholtz Alliance HA216/EMMI, and NSERC.

\end{document}